
\input preprint.sty
\title{Twistors and the BMS Group}
\author{ Adam D. Helfer}
\address{Department of Mathematics, Mathematical Sciences Building, University
of Missouri, Columbia, Missouri 65211, U.S.A.}
\shorttitle{Twistors and the BMS Group}
\pacs{0430, 0420}
\jnl{Classical and Quantum Gravity}

\date

\beginabstract
We compute the motions of null infinity to which the components of the angular
momentum of the gravitational field, as defined by Penrose, are  conjugate.  We
find that the boosts are supplemented by anomalous
translations.  If $c_{ab}$ is the skew bivector determining the component of
the angular momentum in question, the anomaly is proportional to
$$c_{ab}t^b\phi ^{ad}\, ,$$
where $t^a$ is a unit vector in the direction of the Bondi--Sachs momentum and
$\phi _{ab}$ is the quadrupole moment of the shear of the cut at which the
angular momentum is evaluated.  This effect persists in the weak--field limit.
This surprising result is a general consequence of the requirement that angular
momentum be supertranslation--invariant in a quiescent regime, and not some
essentially twistorial peculiarity.
\endabstract

\vfill\eject
\font\tenbm=msbm10
\font\sevenbm=msbm7
\newfam\msyfam
\textfont\msyfam=\tenbm
\scriptfont\msyfam=\sevenbm
\def\Bbb#1{{\fam\msyfam #1}}
\def\R{{\Bbb R}}
\def\Spin{{\Bbb S}}

\def\C{{\Bbb C}}
\def\M{{\Bbb M}}
\def\T{{\Bbb T}}
\def\omicron{{o}}

\def\dang{{\delta _{\rm ang}}}
\def\eth{{\Bbb g}}
\def\varOmega{\mathord{\hbox{\hbox{\raise.182ex\hbox{\fourteenrm o}}%
{\hskip -.67em}%
{\vrule height .3ex width .01em depth -.2ex%
\vrule height .2ex width .02em depth -.1ex%
\vrule height .1ex width .03em depth 0pt%
\vrule width .48em height .40pt depth 0pt%
\vrule height .1ex width .03em depth 0pt%
\vrule height .2ex width .02em depth -.1ex%
\vrule height .3ex width .01em depth -.2ex%
}}}}
\def\unxx#1{\mathop{\vtop{\ialign{##\crcr
      $\hfil\displaystyle{#1}\hfil$\crcr}}}\limits}
\def\unn#1{{\unxx{#1} _{\mathstrut 0}}}
\def\uno#1{{\unxx{#1} _{\mathstrut 1}}}
\def\unt#1{{\unxx{#1} _{\mathstrut 2}}}
\def\unth#1{{\unxx{#1} _{\mathstrut 3}}}
\def\whst{(\widehat M,{\widehat g}_{ab} )}
\def\st{{(M,g_{ab})}}

\def\whSig{{\widehat{\Sigma}}}
\def\cN{{\cal N}}
\def\whG{{\widehat\Gamma}}

\def\scrip{{{\cal I}^+}}
\newcount\footcount
\footcount=0
\def\fnote#1{\global\advance\footcount by1\footnote{${}^{\the\footcount}$}{#1}}
\newcount\EEK
\EEK=0
\def\eek{\global\advance\EEK by 1\eqno(\the\EEK )}
\def\keek{\global\advance\EEK by 1{}(\the\EEK )}
\def\keq{\global\advance\EEK by 1{}\eq(\the\EEK )}

\section{Introduction}

\noindent It would seem desirable for a number of reasons to quantify the
momentum and angular momentum of general--relativistic
systems.\fnote{Throughout
this paper, momentum and angular momentum refer to the relativistic
quantities.}  There is however a fundamental difficulty:  momentum and angular
momentum are usually defined as conserved quantities associated to symmetries,
and a space--time will generically have no symmetries.
Because of this conflict, if a satisfactory general--relativistic theory of
momentum and angular momentum does exist, its foundations will have to be
different from those of the special--relativistic theory.  It would presumably
reduce to the special--relativistic theory in an appropriate limit, but this
might be a rather singular limit.  (There are two, related, reasons to expect
this.  First, the gauge freedom in general relativity is quite different from
that in special relativity; and second, space--times which do admit symmetries
may be expected to be singular points in the phase space of gravitational
degrees of freedom, depending on how this phase space is defined.  Cf.
Moncrieff 1975, 1976.)
The most prominent features of the special--relativistic theory might be wholly
non--existent in a general--relativistic theory, and the key parts of a
general--relativistic theory might be present only indirectly in the
usual formulation of the special--relativisitc theory.

At a minimum, the features one would like in a general--relativistic theory
are positivity of energy, subadditivity (on account of binding energies) of
mass
robustness, covariance, physically acceptable results in test cases, and
breadth
of applicability.  At present, there is no definition which has all these
properties in the most general circumstances, and it may turn out that no such
definition exists.  Still, there is reason to be hopeful.  Good theories of
momentum exist at null (Bondi 1960, Sachs 1962a, Bondi et al 1962) and spatial
(Arnowitt et al 1962) infinity.  Some progress has been made towards the
definition of angular momentum at spatial infinity (Ashtekar and Hansen 1978,
Sommers 1978).  And Penrose's (1982) outline of a twistorial program to define
momentum and angular momentum quasilocally has had enough successes (Shaw 1983;
Tod 1983, 1984, 1986; Helfer 1990) that it seems quite possible \it some \rm
theory will be constructible.  On the other hand, it is now known that the
original program will have to be modified in some respects:  at the least, the
reduction of the symmetry group from the conformal group (which is the twistor
symmetry group) to the Poincar\'e group\fnote{Here the Poincar\'e group is
defined to be the identity--containing component of the isometry group of
Minkowski space.} must be more complicated than had been thought (Helfer 1992,
compare Dougan and Mason 1991).

In trying to extend the definitions of momentum and angular momentum to
general--relativistic systems, it is best to study situations in which one has
a
much control as possible, ideally where only one novel difficulty appears at
each step.  This paper forms part of such a program.  We shall study the
relation between the twistorial definition of angular momentum at null infinity
and the Hamiltonian theory of conserved quantities.  As would be expected from
the arguments above, this relation is less straightforward than the usual
relation between conserved quantities and symmetries.  We shall find that there
\it is \rm a relation, and that the indirectness of this relation goes some way
to resolving the mystery of why apparently reasonable definitions of angular
momentum at null infinity, when based on the Bondi--Metzner--Sachs group (the
asymptotic symmetry group), do not give physically satisfactory results in
general.

The essential idea will be to define a phase space of radiative modes of the
gravitational field, and then invert the usual relation between conserved
quantities and symmetries:  we shall start from the twistorial angular
momentum,
and deduce the motions of the phase space to which they are conjugate.  Since
the phase space is a certain function space, \it a priori \rm these motions
could be complicated non--local funcationals of the data.  Remarkaby, it turns
out that they are induced by certain BMS motions.  These motions are those
which
(in a suitable sense) might be expected, except that the boosts are
supplemented
by ``anomalous'' translations.

We shall discuss the program in more detail as we give a plan of the paper.  In
order to do this, we must first recall some properties of the BMS group.

\subsection{The BMS Group}

\noindent For the space--times usually considered to model isolated systems
with
outgoing gravitational radiation, future null infinity $\scrip$ is
diffeomorphic
to $\{ (u,\theta ,\varphi )\mid u\in\R$, $(\theta ,\varphi )\in S^2\}$.  (The
coordinates are called \it Bondi coordinates, \rm and are determined ---
although not uniquely --- by geomteric considerations.)  The BMS group is
generated by two sorts of motions:  Lorentz motions,\fnote{Here the Lorentz
group is the identity--containing component of $O(1,3)$.} which act in much the
same way as they do on half of a light--cone; and \it supertranslations, \rm
each of which has the form (in Bondi coordinates)
$$u\mapsto u+\alpha (\theta ,\varphi )\, ,\qquad\theta\mapsto\theta\, ,
  \qquad\varphi\mapsto\varphi\, .\eek$$
for some smooth function $\alpha$.  In fact, the BMS group is the semidirect
product of the supertranslations and the Lorentz motions,
$$0\to\hbox{Supertranslations}\to\hbox{BMS}\to\hbox{Lorentz}\to 0\, ,\eek$$
in formal analogy to the way the Poincar\'e group is the semidirect product of
the translations and the Lorentz motions,
$$0\to\hbox{Translations}\to\hbox{Poincar\'e}\to\hbox{Lorentz}\to 0\, .\eek$$
The relation is still closer, because there is a unique four--dimensional
subgroup of Supertranslations which is normal in BMS; this subgroup may be
identified with the translations.  There is no invariant sense to a
``translation--free supertranslation,'' however.

\subsection{Plan of the Paper}

\noindent The next Section covers preliminaries:  we recall some terminology
for
complex vector fields on complex manifolds (the phase space will be a subspace
of a complex manifold); review the definitions of spin-- and boost--weighted
functions, especially those which are ``electric'' and ``magnetic;'' and derive
a formula for the action of the Lorentz group on these functions.

In Section 3, the definition of the phase space is presented.  The main idea is
to develop a treatment in which the outgoing radiative and internal modes of
the gravitational field are as nearly as possible decoupled, for we are
interested in the analysis of radiation.  The phase space is a refinement of
one
proposed earlier by Ashtekar and Streubel (1981).  (These authors also seem to
be the first to have tried to develop such a phase space.  A hint of the idea
may be present in Sachs's 1962b ``classical commutation relations at
infinity.'')  The phase space was rigorously constructed, using recent theorems
of Friedrich (1983) and Rendall (1990, 1992) in Helfer (1993), and in this
section we restate the main results of that paper.

Section 4 treats the reduction of the phase space.  It is here that we modify
the usual treatment of the asymptotic symmetries.  In outline, the argument is
this:  The phase space, as constructed, contains spurious, ``gauge,'' degrees
of freedom.  These are the relics, at null infinity, of general coordinate
invariance:  they are the diffeomorphisms which preserve the asymptotic
structure.  However, not \it all \rm BMS motions should be regarded as gauge in
this sense.  (If for example we were to reduce by the translations, for
example, there would be no possibility of identifying the momentum at different
values of the Bondi parameter $u$.)  Somehow it is necessary to elimate
``most''
of the BMS motions, but be left with the Poincar\'e group.  We shall show that
there is an action by the quotient group Supertranslations/Trans\-la\-tions on
the phase space; the freedom that remains after this is ``the same size'' as
the
Poincar\'e group.

The action is easy to write down but requires some care to interpret.  The
phase space will essentially consist of the different possible Bondi shears
$\sigma$ on $\scrip$.  If we fix a Bondi system, and consider the shears as
functions on the abstract manifold $\scrip$ (rather than on its image $\R\times
S^2$ in coordinates), the action is
$$\sigma\mapsto\sigma +\eth ^2\alpha\, .\eek$$
This action is \it not quite \rm the standard action of the BMS group by
diffeomorphisms, except in stationary space--times.  (The standard action
would be $\sigma\mapsto\sigma\circ \phi ^{-1}+\eth ^2\alpha$, where $\phi$ is
th
diffeomorphism induced by the supertranslation.  It is also not the same as a
passive BMS motion, which would have the form above, but be accompanied by a
change in coordinates).  Indeed, precisely because there is no preferred
complement of Translations in Supertranslations, we cannot expect to reduce by
a
standard action by diffeomorphisms.

The details of the reduction of the phase space largely follow the
procedure of Bergmann and Goldberg (1955) and Dirac (1964).  Let us recall that
there are two elements to such reductions:  formation of the quotient by a
symmetry; and restriction to the constraint surface.  In our case, it will be
passage to the constraint surface which is most important.  This will restrict
the admissible tangent vectors in the phase space, and such a restriction turns
out to be essential in order to solve for the Hamiltonian vector field
conjugate to the angular momentum.  There will be no reason to restrict to the
zero surface of the constraints, but we may consider other level surfaces as
well.  The level surfaces are precisely the infrared sectors.
Because the action is not induced by diffeomorphisms, it is not desirable to
pass to the quotient of the phase space by the action.  Instead, we shall use a
natural slice for the action.

Section 5 begins the twistorial construction.  Penrose's procedure associates
a twistor space $\T (S)$  with certain structure to each cut $S$ of $\scrip$,
an
from this is constructed a Minkowski space $\M (S)$ of origins; the angular
momentum can be thought of as ``living on'' this Minkowski space.  Since the
entire construction depends on the data for the gravitational field at $S$,
which vary with the point in the phase space, what we really have is (for each
fixed $S$) a bundle of twistor spaces over phase space, and so a bundle of
Minkowski spaces over phase space.  In order to meaningfully differentiate the
angular momentum as a function on phase space (so as to apply the Hamiltonian
formalism), it will be necessary to define a connection on this bundle.
It turns out that there is a natural connection on this bundle, which preserves
all of the relevant structure, and is integrable.  This comes about because
each
Minkowski space $\M (S)$ has a natural origin (Helfer 1990).

In Section 5, we also show that the twistorial angular momentum is
cut--independent, in a suitable sense, in a quiescent regime.

In Section 6, we compute the Hamiltonian vector fields conjugate to the total
angular momentum of the space--time, in the sense of the Bondi retarded time
parameter $u$ tending to $-\infty$.  In Section 7, we interpret the terms of
the vector fields.

Section 8 contains a summary and some discussion.

\subsection{The System Considered}

\noindent The idea of inverting the usual relation between symmetries and
conserved quantities is evidently rather general.  In the present paper, it is
applied to discover the motions of the gravitational radiation field to which
the twistorial angular momenta are conjugate, in the case that no matter fields
are present at null infinity.  Further applications will be given elsewhere.

The physical system we have in mind is a space--time which is quiescent (as
far as gravitational radiation goes) except for an interval between two
retarded times.  For analytic simplicity, we are going to assume that in the
quiescent periods the Bondi news function vanishes.  This is a mathematical
idealization which would not be expected to be valid generically.  (It will
follow from the construction of the phase space, though, that a broad class of
solutions does have this behavior.)  Its main force in the present paper is to
give a clean isolation of the radiative from the internal modes; one could
pursue a similar program, taking into account internal modes, for systems in
which ``tail'' effects are expected to be significant.  This will be done
elsewhere.

There is some question about exactly what one should demand for quiescence:
one wants the shear to be $u$--independent; ought it also be purely electric?
On the basis of linear theory, one would expect the magnetic part of $\sigma$
to vanish as $u\to\pm\infty$ for a large class of systems (Newman and Penrose
1966), and usually one imposes this condition.  Also, the magnetic part of the
shear will be zero in a stationary space--time, so if we are considering
space--times which are close to stationary before the emission of radiation, it
is natural to require the magnetic part of $\sigma$ to vanish as $u\to
-\infty$.  We shall for simplicity require the shear to be purely electric
before and after the emission of radiation.  It is worth pointing out that
there is a time--asymmetry in the \it naturality \rm of this condition.  For
outgoing radiation problems, it is natural because one has in mind a system
that is initially and finally close to stationary:  one restricts one's
attention to such systems.  But for incoming radiation problems, the condition
amounts to a coherence requirement on the radiation put into the system, which
is less natural.

The angular momentum we are concerned with here will be the total angular
momentum of the space--time, as measured at future null infinity.  This may be
defined as the angular momentum at sufficiently early retarded times (before
any
gravitational waves have been emitted).  It is not known at present how this is
related to the total angular momentum as measured at spacelike infinity; in
fact, it is not known if the space--times we consider admit spatial infinties
with the sorts of asymptotics for which definitions of the angular momenta are
known.  However, it seems quite possible that in the event the angular momentum
at spatial infinity can be defined, it will be equal to the quantity we
study.  (The corresponding equality for momenta was established by Ashtekar
and Magnon--Ashtekar 1979.)

It is also possible to treat the angular momentum at a finite retarded time by
these techniques.  The results are similar, but there are some additional
conceptual issues, and this will be done elsewhere.

We have chosen to work in the $C^\infty$ category because the theorems of
Friedrich and Rendall allow it, and in the absence of any other choice it seems
the most natural.  However, essentially no change would be expected for fields
of finite differentiability as long as the leading terms of the curvature
components $\Psi _4$, $\Psi _3$, $\Psi _2$ and $\Psi _1$ peel properly.

\it Conventions.  \rm
We will be concerned with calculus on the phase space and on the
space--time.  It will be convenient to use the standard spin--coefficient
formalism on the space--time (Penrose and Rindler 1984--6), and the modern
coordinate--free notation for the phase space.  We use $\langle
s,\lambda\rangle$ to stand for the duality pairing between a vector $s$ and a
covector $\lambda$ in this case.  Thus the symbol $\d$ represents a gradient in
the phase space except when it occurs in a line, area or volume element, $\d
u$,
$\d{\cal S}$ or $\d{\cal S}\, \d u$, with respect to the Bondi coordinate
system
on $\scrip$.  The symbol $\Omega$ stands for the conformal factor and
$\varOmega$ for the symplectic form.

An introduction to phase spaces for field theories, as infinite--dimensional
sym\-plectic manifolds, will be found in Chernoff and Marsden (1974).

\section{Preliminaries}

\noindent We collect here some definitions and terminology which will be used
in
the rest of the paper.  Also we derive a formula, which seems not to have been
given previously, for the action of the Lorentz group on spin-- and
boost--weighted functions.

\subsection{Complex Vector Fields}

\noindent The phase space we shall work with will be a subspace of a complex
vector space.  We shall use this complex structure, although at an elementary
level, and we want here to recall the definitions of $(1,0)$ and $(0,1)$ vector
fields.

In local coordinates, a complex vector field on
a complex manifold has the form $v^a\partial /\partial z^a +{\tilde
v}^{\overlin
a}\partial /\partial {\overline z}^{\overline a}$.  (Here $z^a$ are local
holomorphic coordinates.)  The first term is the \it $(1,0)$ part \rm of the
vector field, and the second term is the \it $(0,1)$ part.  \rm

A more abstract definition, suitable for generalization to
infinite--dimensional spaces, is this.  Let $W$ be a complex vector space.
(The space $W$ will be the tangent space to a point in a complex manifold.)
Then $W$ has an underlying real vector space $W_\R$ (got by restricting
multiplication by scalars to real scalars), and a \it complex vector \rm in $W$
is defined to be an element in the complexification of $W_\R$.  This
complexification is $W_\R\otimes\C$ which may be canonically identified with
$W\oplus\overline{W}$ (where $\overline W$ is the conjugate vector space to
$W$)
by identifying $v+iw\in W_\R\otimes\C$ with $(v+iw,v-iw)\in W\oplus\overline
W$.  We write $W_\R\otimes\C =W^{(1,0)}\oplus W^{(0,1)}$ for this
identification, so $W^{(1,0)}=W$ and $W^{(0,1)}=\overline W$.

This will enter into our analysis as follows.  The phase space will be a
certain function space, of allowable Bondi shears.  Since the Bondi shears take
values in a complex line--bundle, the phase space will be (a subspace of) an
infinite--dimensional complex manifold.  A skew bitensor $c_{ab}$ will
determine a vector field
$$c^{ab}V_{ab}=c^{A'B'}V_{A'B'}+{\overline c}^{AB}{\overline V}_{AB}\eek$$
on phase space.  Here the left--hand side is insensitive to the complex
structure of the phase space (it can be thought of as an element in $W_\R$).
Each term on the right--hand side must be thought of as lying in the
complexification of the tangent space to the phase space, so each of
$V_{A'B'}$, ${\overline V}_{AB}$ has $(1,0)$ and $(0,1)$ parts,
satisfying
$$V_{AB}^{(1,0)}=\overline{V_{A'B'}^{(0,1)}}\qquad\hbox{and}\qquad
  V_{AB}^{(0,1)}=\overline{V_{A'B'}^{(1,0)}}\, .\eek$$
In the twistor formalism, it is most natural to compute $V_{A'B'}$.  The
$(1,0)$ and $(0,1)$ parts of this are identified as complex--linear and
complex--antilinear perturbations of the shear.

\subsection{Spin-- and Boost--Weighted Functions}

\noindent For a full treatment of the spin-coefficient formalism, see Penrose
and Rindler (1984--6).  Here our concern is to fix some conventions and recall
the main formulae.

Let $\Spin ^A$ be spin space.  It is a two--complex dimensional
vector space, and its associated projective space is identified with the
Riemann sphere $S^2$ equipped with a conformal structure and an orientation,
but not a metric.  To accord with standard conventions, we shall use
$\omicron ^A$ for a general (non--zero) element of $\Spin ^A$; the class of
spinors proportional to $\omicron ^A$ defines a point on the sphere.  We may
sometimes introduce a unit future--pointing timelike vector $t^a$.  The
subgroup
of $SL(2,\C )$ which preserves this vector is $SU(2)$ (in a basis in which the
spinor $t^{AA'}$ is diagonal), and so $t^a$ defines a unit sphere metric on
$S^2$.  In this case, we set $\iota ^A=\sqrt{2} t^{AA'}\omicron _{A'}$; then
$\omicron _A\iota ^A=1$.

A function on $\Spin ^A$ which may not be holomorphic is conventionally written
$f(\omicron ^A,{\overline\omicron}^{A'})$.  The function is said to be \it of
type $\{ p,q\}$ \rm if
$$f(\lambda\omicron ^A,{\overline\lambda}{\overline\omicron}^{A'})
  ={\strut\lambda}^p{\strut\overline\lambda}^qf(\omicron
^A,{\overline\omicron}^{A'})\, .\eek$$
(One says $f$ has \it spin--weight \rm $(p-q)/2$ and \it boost--weight
\rm $(p+q)/2$.)  Such a function is a section of a certain complex line--bundle
over the sphere.  The line--bundle will be denoted simply as $\{ p,q\}$.
Its space of smooth sections will be denoted $C^\infty (S^2,\{ p,q\} )$.

There are two important differential operators on these line bundles:
$$\eth :C^\infty (S^2,\{ p,q\} )\to C^\infty (S^2,\{ p+1,q-1\})\eek$$
defined by
$$\eth f=\sqrt{2}t_{AA'}\omicron ^A{\overline\omicron}^{A'}
 \overline{\iota}^{B'}{\partial\over{\partial {\overline\omicron}^{B'}}}f\eek$$
and its conjugate $\eth ':C^\infty (S^2,\{ p,q\})\to C^\infty (S^2,\{
p-1,q+1\})$.

In this paper, the sphere in question will be a cut of $\scrip$, with the
spinor
$\omicron ^A$ one of the elements of the dyad of the Bondi system.  The complex
structure of a cut is preserved by flowing along the vector field $n^a$, so we
may regard the bundle $\{ p,q\}$ on any cut as the pull--back of a bundle
(denoted by the same symbol) on $\scrip$.

\subsection{Action of the Lorentz Group}

\noindent We shall later be concerned with the transformation of certain
quantitites under the Lorentz group which is the quotient
BMS/Super\-trans\-la\-tions.

The group $SL(2,\C )$ acts on the line bundles $\{ p,q\}$ so as to
preserve fibres.  (One says the line bundles are \it homogeneous.\rm )  Let
$\lambda ^A{}_B$ be an element of the Lie algebra.  Then the associated vector
field on $\Spin ^A$ is
$$\lambda ^{AB}\omicron _A{{\partial}\over{\partial\omicron ^B}}\,
.\eek$$\xdef\veceq{{\the\EEK}}
Because this is homogeous of degree zero, it defines an action on each $\{
p,q\}$.  (The action on the Riemann sphere itself can be thought of as the case
$p=q=0$.)  We shall want a formula for this in terms of $\eth$.  On a function
with values in $\{ p,q\}$, we have
$$\eqalign{\lambda ^{AB}\omicron _A{{\partial}\over{\partial\omicron ^B}}
  &=\bigl( \iota ^B\omicron _C-\omicron ^B\iota _C\bigr)
    \lambda ^{AC}\omicron _A{{\partial}\over{\partial\omicron ^B}}\cr
  &=\bigl( \sqrt{2} t\cdot l\bigr) ^{-1}\lambda _0\eth '
    -p\lambda _1\cr
  &=( \sqrt{2} t\cdot l) ^{-1}\bigl( \lambda _0\eth '
    -p(1/2)(\eth '\lambda _0)\bigr) \cr}\eek$$
where we have written $\lambda _0=\lambda _{AB}\omicron ^A\omicron ^B$,
$\lambda _1=\lambda_{AB}\omicron ^A\iota ^B$ as usual.

Now let us consider quantities of type $\{ p,q\}$ which may also be
conformally weighted.  Then the action above will be supplemented by a term
which is minus the conformal weight times the divergence of the vector field on
the sphere, or rather times the $\lambda _{AB}$--linear part of the divergence:
$$( \sqrt{2} t\cdot l) ^{-1}\bigl( \lambda _0\eth '
    -(1/2)(2w+p)(\eth '\lambda _0)\bigr)\, .\eek$$
In fact, the twistor conventions will lead us naturally to consider the
conjugate of this:
$$( \sqrt{2} t\cdot l) ^{-1}\bigl( {\overline\lambda}_{0'}\eth
    -(1/2)(2w+q)(\eth {\overline\lambda}_{0'})\bigr)\, .\eek$$

The case of interest for us below will be a function of the same type and
weight as the Bondi shear (and derived from it).  Let us say $\varsigma$ has
type $\{ 3,-1\}$ and conformal weight $-1$.  Then the action is
$$c\eth \varsigma
    +(3/2)(\eth c)\varsigma\, ,\eek$$
where $c={\overline\lambda}_{0'}$,
in a Bondi frame, and the action on $\overline\varsigma$ is
$$ c\eth \overline\varsigma
-(1/2)(\eth c)\overline\varsigma \, .\eek$$

\subsection{Electric and Magnetic Functions}

\noindent For the material here, see Newman and Penrose (1966).

Since the conjugate of $\{ p,q\}$ is $\{ q,p\}$, spin--weight zero quantities
(those with $p=q$) have unique $SL(2,\C )$--invariant decompositions into real
and imaginary parts.  There is also a hidden $SL(2,\C )$--invariant Hermtian
structure on certain of the other line bundles under consideration.  It turns
out that the operator $\eth ^{2w+q+1}$ (for $2w+q+1$ a positive integer) is
$SL(2,\C )$--invariant on type $\{ p,q\}$ functions of conformal weight $w$.
Let us consider the case $p=q$.  Then the operator is surjective, and its
kernel
is the complexification of a real $SL(2,\C )$--invariant subspace.  This allows
us to write each element in the range as the sum of the image of a real part
and
the image of an imaginary part, these images being uniquely defined.  These are
called the \it electric \rm and \it magnetic \rm parts of the element.

The important case for us will be functions of the type of the Bondi shear:
$\{
3,-1\}$ and conformal weight $-1$.  Each such function $\varsigma$ can be
written as $\eth ^2\phi$ for some $\phi$ of type $\{ 1,1\}$ and conformal
weight $+1$.  The electric part of $\varsigma$ is $\eth ^2\Re\phi$ and the
magnetic part is $\i\eth ^2\Im\phi$.

\section{The Phase Space}

\noindent The phase space of radiative degrees of freedom was constructed in
Helfer (1993), building on work of Ashtekar and Streubel (1981), Friedrich
(1983) and Rendall (1990, 1992).
Here we shall summarize the results.

We are concerned with a space--time $\whst$ which has the following
properties.  It is oriented and time--oriented, the manifold ${\widehat M}$
embeds as the interior of a manifold with boundary $M$, and

\itemitem{} (a) there is a smooth function $\Omega$ on $M$, positive on
${\widehat M}$ and zero on $\scrip =\partial M$, with non--zero gradient on
$\scrip$;

\itemitem{} (b) $g_{ab}=\Omega ^2{\widehat g}_{ab}$ extends to a smooth
non--degenerate metric on $M$;

\itemitem{} (c) $\scrip$ is a $g_{ab}$--null hypersurface, diffeomorphic to
$\R\times S^2$, with the ``$\R$'' factors being the null generators, and every
point on $\scrip$ is the future end--point of a null geodesic in $\whst$;

\itemitem{} (d) there is a partial Cauchy surface $\whSig$ in $\whst$ whose
closure $\Sigma$ in $\st$ is compact and meets $\scrip$ transversely in a cut
$Z$;

\itemitem{} (e) the generators of $\scrip$ are complete to the past in a Bondi
frame.

One can use the results of Friedrich and Rendall to show
that such space--times are determined by two sorts of data:  certain Cauchy
data
on $\whSig$; and asymptotic characteristic data on $\cN$, the part of $\scrip$
t
the past of $Z$.  We shall call the data on $\whSig$ the \it final state, \rm
an
the asymptotic characteristic data on $\cN$ the \it radiation data.  \rm

The manifold underlying the phase space ought to be a function space built out
of the final states and the radiation data.  Since the radiation data must
match
up to the final state at $Z$, the phase space should have a bundle structure,
with the radiative degrees of freedom fibering over the final states.  Here we
shall be concerned only with a fibrewise analysis.  Fix, then, a final state.

{}From the data on $\whSig$, we can work out (once a Bondi frame has been
chosen)
the Bondi shear at $Z$, and all its $u$--derivatives there.  For simplicity, we
are going to assume that $\sigma$ is electric and independent of $u$ near $Z$
an
near $u=-\infty$.
It will be convenient in
what follows to fix a Bondi system in which $Z$ is given by $u=0$.

Now let $\cN =\{ (u,\theta ,\varphi)\mid u\leq 0$, $(\theta ,\varphi )\in
S^2\}$
and let $\whG$ be the space of all smooth functions of type $\{ 3,-1\}$ on
$\cN$ which are electric and independent of $u$ in some neighborhood of
$u=0$ and in some neighborhood of $u=-\infty$.  The space
$\whG$ is somewhat larger than the phase space we are ultimately interested in,
since we are allowing different shears at $u=0$ in $\whG$.  This is to
accomodate the reduction which will be discussed in the next Section.

The is a natural topology and smooth structure on $\whG$, which may
be characterized as follows.  Choose any function $f$ (of type $\{ 0,0\}$) on
$\cN$ which is smooth, identically $-1$ for sufficiently negative $u$ and
identically $+1$ for sufficiently large $u$.  The map
$$C^\infty (S^2,\{ 3,-1\} )_{\rm electric}\oplus C^\infty (S^2,\{ 3,-1\} )_{\rm
electric}
  \oplus C_0^\infty (\cN ,\{3,-1\})  \to
\Gamma\eek$$\xdef\charry{\the\EEK}%
given by
$$(\alpha ,\beta ,\gamma )\mapsto \alpha f+\beta  +\gamma\eek$$
is an isomorphism of (algebraic) vector spaces.  (Here the subscript $0$
denotes compact supports.)  Topologize the right--hand side so that this is a
homeomorphism.  This is independent of $f$ and BMS--invariant in
the appropriate senses.  The inverses of these maps, as $f$ varies, define a
smooth atlas on $\whG$.

The symplectic form is given by
$$\widehat{\varOmega} (s_1,s_2)=(8\pi G)^{-1}\int _{\cN}\bigl[ {\dot
s}_1\overline{s_2}-
   \dot{\overline{s}}_2 s_1\bigr] \, \d u\d{\cal S}
  +\hbox{ conjugate}\, .\eek$$
It is smooth, closed and weakly non--degenerate.  It agrees with a suitable
limit of symplectic forms of the $3+1$ formalism.

It is not known whether to each point in $\whG$ there corresponds a space--time
satisfying condition (e), that is, having a future null infinity which contains
all of $\cN$.  It is possible that only certain data give rise to such
space--times.  However, this is no limitation for the present paper, since we
are concerned only with local analysis on the phase space, and the stability
results of Rendall (1992) guarantee that the subspace of $\whG$ for which (e)
holds is open.  In what follows, we shall implicitly assume we are working in
this open set.

There is a convention in the way we have set up the phase space which
is worth discussing here, as it will be important in what follows.
The function $\sigma$ is supposed to measure the shear of the $u=$ constant
cuts of the Bondi coordinate system.  Clearly, this interpretation of $\sigma$
depends on the Bondi system chosen.  For our phase space, the Bondi system has
been fixed, and we consider different possible functions $\sigma$.  Each such
function, by the theorems of Friedrich and Rendall, determines a space--time
for which $\sigma$ has the interpretation of the Bondi shear.
The most important consequence of this convention is that the spinor dyad, and
hence the operators $\eth$ and $\eth '$, are independent of $\sigma$.

\section{Gauge Freedom}

\noindent In Hamiltonian approaches to General Relativity, one usually
constructs a phase space which has spurious, ``gauge,'' degrees of freedom.
This space must be reduced to isolate the physical degrees of freedom.
Something like this will happen for our phase space, but there will be some
important differences between this and the usual reduction.  In order to
explain these, it will be necessary to recall the Bergmann--Dirac theory in the
formalism of symplectic manifolds.
In this sketch, we ignore all technical difficulties (whether quotients are
manifolds, problems in infinite dimensions, etc.).

Suppose a phase space $(\Gamma ,\varOmega )$ is given, together with a
first--class family $\{ C_\alpha\mid \alpha\in A\}$ of constraints.\fnote{A
family of constraints is \it first--class \rm if it is closed under Poisson
brackets (Bergmann and Goldberg 1955, Dirac 1964).}  The restriction of the
symplectic form to the constraint manifold $\Gamma _0=\{ p\in\Gamma\mid
C_\alpha
(p)=0\}$ is degenerate, and this degeneracy is compensated by introducing an
equivalence relation.  Let $V_\alpha$ be the Hamiltonian vector field for
$C_\alpha$:
$$\d C_\alpha =\varOmega (\cdot ,V_\alpha )\, ,\eek$$\xdef\funeq{\the\EEK}%
and ${\cal V}$ be the Lie algebra of these vector fields.
Then two points in $\Gamma _0$ are identified if there is a vector field in
${\cal V}$ whose flow takes the first to the second point.

We may run this backwards.\fnote{Jan Segert points out to me this is the
reduction of Marsden and Weinstein (1974).}  Suppose a connected, abelian (for
simplicity) Lie group acts on a phase space $(\Gamma ,\varOmega )$.  (So the
action gives a representation of the Lie group by symplectomorphisms.)  Let $A$
be the Lie algebra of the group, and ${\cal V}=\{ V_\alpha\mid \alpha\in A\}$
the Lie algebra of the vector fields on $\Gamma$ generated by the action.
Suppose further that we can find a \it moment map \rm $C:\Gamma\to A^*$ so that
equation (\funeq ) is satisfied.  Then to form a new phase space by reducing by
the action of the Lie group, we should also, in order to have a non--degenerate
symplectic form, restrict to a constraint surface.  One need not restrict to
$\Gamma _0$,  though; one could restrict to $\Gamma _c=C^{-1}(c)$ for some
$c\in
A^*$.  From this point of view, the original phase space is foliated by the
$C=$ constant subspaces, and reduction includes a restriction to such a
subspace.

Notice that there are two sorts of foliation involved in the reduction:  that
just discussed, by the level surfaces of the moment map; and the foliation by
the orbits of the action of the gauge group.  \it We shall always use the terms
\rm leaf \it and \rm foliation \it to refer to the former, a level surface
of the moment map and the family of level surfaces.  \rm  Notice that we
may speak of this foliation whether or not we have passed to the quotient by
the group action.  In what follows, it will be such a foliation which is
crucial.

Now we explain how this applies to the phase space of radiative modes.

It is well--known that the BMS group is the symmetry group of null infinity, in
the sense that the universal structure of null infinity (i.e., that which is
independent of the particular asymptotically flat space--time to which null
infinity is adjoined) is preserved by the BMS group.\fnote{Here we must include
in the definition of ``asymptotically flat'' the requirement that the
generators of $\scrip$ be infinitely long, or else, strictly speaking, the BMS
group does not act on $\scrip$.  In the absence of this requirement, the
discussion applies at the Lie algebra level.}  The BMS group is large because
this universal structure is weak.  ``Most'' of the degrees of freedom in the
BMS
group should be thought of as compensating for lack of structure, rather than
preserving dynamically interesting structures.

We do \it not \rm want to reduce by the Poincar\'e group, since we want to be
left with enough degrees of freedom to identify the momentum and angular
momentum at different cuts.  Roughly speaking, then, we wish to reduce by those
BMS motions which are not Poincar\'e motions, that is, by the
``translation--free supertranslations.''  There is no invariant sense to this,
however.  This means that we cannot expect to pass to a quotient of our phase
space by such an identification.  (In other words, we are not looking for an
action which will identify isometric space--times.)

There is a way around this.  There is an action of the quotient group
Super\-trans\-lations/Trans\-la\-tions on the phase space.  It is given by
$$\sigma\mapsto\sigma +\eth ^2\alpha\, .\eek$$
(Since the translations are precisely the supertranslations annihilated by
$\eth ^2$, it is clear that this passes down to the quotient.)  Some care is
necessary in interpreting this action.  With our conventions, the shear is
thought of as a function on the abstract manifold $\scrip$, so this action is
formally the same as that induced by a passive BMS motion.  However, we do \it
not \rm accompany this by a change in the Bondi coordinate system.  Note that
this is also \it not \rm the same as an active BMS motion; such a motion would
rather be
$$\sigma\mapsto\sigma\circ\phi ^{-1}+\eth ^2\alpha\, ,\eek$$
where $\phi :\scrip \to\scrip$ is the diffeomorphism induced by $\alpha$.

The foliation will be important, because the admissible tangent vectors will be
those tangent to the leaves.
Let $V_\alpha$ be the vector field on $\Gamma$ inducing this action.  We have
$$\eqalignno{\varOmega (s,\eth ^2\alpha )&=
    (8\pi G)^{-1}\int {\dot s}{\eth '}^2\alpha\, \d u\d{\cal S} +\hbox{
conjugate}\cr
  &=(8\pi G)^{-1}\oint [\! [ s]\! ]{\eth '}^2\alpha\, \d{\cal S}+\hbox{
conjugate}\cr
&=\langle s,\d (8\pi G)^{-1}
  \oint [\! [\sigma ]\! ]{\eth '}^2\alpha\, \d{\cal S}+\hbox{
conjugate}\rangle\, ,&\keek\cr}$$
where $[\! [f]\! ] =f(0)-f(-\infty )$.  Therefore the associated constant of
motion is
$$C_\alpha =(8\pi G)^{-1}\oint
  [\! [\sigma ]\! ]{\eth '}^2\alpha+\hbox{ conjugate}\, .\eek$$
As $\alpha$ varies, these constants detect $[\! [\sigma
]\! ]$.
Two shears are said to lie in the same \it infrared sector \rm if they have the
same values of $[\! [\sigma]\! ]$.  Thus the leaves of the phase space
are precisely the infrared sectors.

It is undesirable to pass to the quotient of $\whG$ by this action, because
this action is \it not \rm that induced by a diffeomorphism and so does not
take one radiation datum into an equivalent --- in the sense of belonging to an
isometric space--time --- one.  However, it is natural to fix a slice of the
action by considering only those $\sigma$'s which agree at $Z$ with that
deduced by the final state.  This we shall do.  Then each vector tangent to a
leaf has a canonical representative $s$ which is zero near $Z$ and near
$u=-\infty$.

It is straightforward to verify the technical aspects of the reduction:  that
each leaf (intersected with the slice) is a manifold, that $\varOmega$ is
smooth and weakly non--degenerate on each.  We denote by $\Gamma$ one
such intersection.  Thus the elements of $\Gamma$ all agree with the shear
induced by some final state at $Z$ and at $u=-\infty$.

\section{The Bundle of Twistor Spaces and its Connection}

\noindent To compute the momentum and angular momentum at a cut $S$, the
twistor program first constructs a twistor space $\T (S)$, then certain
additional structure on $\T (S)$, and from these a Minkowski space
$\M (S)$.  The momentum and angular momentum naturally refer to this Minkowski
space.  The Minkowski spaces for the different cuts are canonically
identified modulo translations, and this is enough to compare the momenta at
different cuts.  In general, there is no known canonical identification without
factoring out by the translations, and this means that it is not
straightforward
to compare angular momenta at different cuts.  One might expect it to be
equally difficult to compare angular momenta as the data for the field are
altered.  However, it turns out in this case that there \it is \rm a natural
identification of the Minkowksi spaces.

\subsection{The Connection}

\noindent Let us recall the construction of $\T (S)$.  It is the space of
solutions $(\omega ^0,\omega ^1)\in C^\infty (S,\{ -1,0\} )\times C^\infty
(S,\{
1,0\} )$ of the two--surface twistor equations on $S$,
$$\left.\eqalign{\eth '\omega ^0&=0\cr\eth\omega ^1&=\sigma\omega ^0\cr}
  \right\}\, .\eek$$\xdef\twoeq{{\the\EEK}}%
It is known that $\T (S)$ is four--complex dimensional.  Indeed, it is possible
to construct it explicitly.  Let $\phi$ by the solution to $\eth ^2\phi
=\sigma$ which has zero $l=0$ and $l=1$ components with respect to a metric on
$S$ determined by the Bondi--Sachs momentum.  Then the solutions to (\twoeq{}a)
form a two--dimensional family independent of $\sigma$ (and are in fact the
spin--weight $-1/2$ spherical harmonics with $j=1/2$).  The solutions to
(\twoeq{}b) are the functions of the form
$$\omega ^1=\omega ^0\eth\phi -\phi\eth\omega ^0 +\xi\,
,\eek$$\xdef\xisub{{\the\EEK}}%
where $\xi$ has type $\{ 1,0\}$ and satisfies
$$\eth \xi =0\, .\eek$$
(That is, $\xi$ is a spin--weight $+1/2$ spherical harmonic with $j=1/2$.  The
equation $\eth ^2\omega ^0=0$ is a consequence of $\eth '\omega ^0=0$.)  The
function $\phi$ is called the \it focus \rm of the cut (Helfer 1990).  We may
think of $\phi$ as given by
$$\phi =\eth ^{-2}\sigma\, ,\eek$$
where $\eth ^{-2}$ represents the required Green's operator.  Since the
kernel of this operator depends smoothly on the Bondi--Sachs momentum at $S$,
which in turn depends smoothly on the Bondi shear on $\cN$, the focus is a
smooth function on $\Gamma$.  It follows that the family of twistor spaces $\T
(S)$ as $\sigma$ varies is naturally a smooth bundle over $\Gamma$.

We define a connection $\nabla$ on this bundle by
$$\left.\eqalign{\nabla\omega ^0&=\d\omega ^0\cr
  \nabla\omega ^1&=\d\omega ^1-\omega ^0\eth \d\phi +(\d\phi )\eth\omega ^0\cr}
  \right\}\, .\eek$$
(Recall our convention is that $\d$ is the exterior derivative on $\Gamma$.)
It is easily verified that this \it is \rm a connection.  Notice that in terms
of the substitution (\xisub{}), we have
$$\left.\eqalign{\nabla\omega ^0&=\d\omega ^0\cr
  \nabla\xi&=\d\xi\cr}
  \right\}\, .\eek$$
Therefore the curvature of the connection is zero, and the isomorphism $(\omega
^0,\omega ^1 )\mapsto (\omega ^0,\xi )$ provides a trivialization of the
bundle.

As noted above, the Minkowski space $\M (S)$ is recovered from the vector space
$\T (S)$ and some additional structures:  an infinity twistor, an alternating
twistor, and a pseudo--Hermitian inner product.  The group preserving these
structures is locally isomorphic to the Poincar\'e group, and so $\M (S)$ is
constructed as an affine space.  In fact, the focus defines an origin twistor
too, so there is a natural origin in $\M (S)$, which one can think of as a sort
of average focus of the null hypersurface ingoing from $S$ in the interior of
the space--time.  The connection has been contrived to preserve all these
structures, so the Minkowski space $\M (S)$ is independent of the shear.
(More formally, the bundle of Minkowski spaces is canonically trivialized.)

In order to show that the twistor structures are preserved, we shall give the
expressions for all of them in terms of the pairs $(\omega ^0,\xi)$.  All of
these may be easily verified.  We write $Z^\alpha$ for such a pair, thought of
as defining an abstract element of $\T (S)$.

The infinity twistor is given by
$$I_{\alpha\beta}{\uno Z}^\alpha{\unt Z}^\beta
  ={\unt\omega}^0\eth{\uno\omega}^0-{\uno\omega}^0\eth{\unt\omega}^1\, ,\eek$$
the alternating twistor by
$$\epsilon _{\alpha\beta\gamma\delta}
  {\unn Z}^\alpha {\uno Z}^\beta {\unt Z}^\gamma {\unth Z}^\delta
  =\left|\matrix{
  {\unn\omega}^0 &{\uno\omega}^0 &{\unt\omega}^0 &{\unth\omega}^0\cr
  {\unn\xi} &{\uno\xi} &{\unt\xi} &{\unth\xi}\cr
  \eth{\unn\omega}^0 &\eth{\uno\omega}^0 &\eth{\unt\omega}^0
     &\eth{\unth\omega}^0\cr
  \eth '{\unn\xi} &\eth '{\uno\xi} &\eth '{\unt\xi} &\eth '{\unth\xi}\cr
}\right|\, ,\eek$$
the norm by
$$H_{\alpha{\overline\alpha}}Z^\alpha {\overline Z}^{\overline\alpha}
  =\i\bigl( {\overline\omega}^{0'}\eth '\xi -\xi\eth '{\overline\omega}^{0'}
    \bigr) +\hbox{ conjugate }\, .\eek$$
The origin twistor $O_{\alpha\beta}$ is defined up to scale by the condition
$O_{\alpha\beta}Z^\beta =0$ if and only if $\xi =0$.  (A scale may be fixed,
too, but we shall not need one.)  Since all of these formulas are independent
of $\sigma$, the preservation of the structures by the connection is evident.

\subsection{Computation of $\d\phi$}

\noindent We shall need an explicit formula for $\d\phi$.

The focus $\phi$ of the cut $S$ is defined to be the solution of $\eth ^2\phi
=\sigma$ with zero $l=0$ and $l=1$ parts with respect to the frame defined by
the Bondi--Sachs momentum at $S$.  Let us write $\eth ^{-2}$ for the Green's
operator which produces a type $\{ 1,1\}$ field with zero $l=0$ and $l=1$ parts
relative to the Bondi--Sachs momentum.  Also let us put $\delta\phi =\langle
s,\d\phi\rangle$, and similarly for other quantities which change with the
shear.  Then we have
$$\delta\phi =\eth ^{-2}s+\xi\, ,\eek$$
where $\xi$ has only $l=0$ and $l=1$ parts and is determined by the requirement
that the perturbed focus have zero $l=0$ and $l=1$ parts with respect to the
perturbed Bondi--Sachs momentum.  This condition is
$$\langle s,\d\oint n_a\phi (t\cdot n)^{-2}\d{\cal S}\rangle =0\, ,\eek$$
that is
$$\oint n_a\delta\phi\, (t\cdot n)^{-2}\d{\cal S}
   -2\oint n_a\phi (\delta t\cdot n)(t\cdot n)^{-3}\d{\cal S}=0\,
,\eek$$\xdef\anot{\the\EEK}%
where $t^a$ is the unit vector parallel to the Bondi--Sachs momentum.
This is essentially a condition relating spherical harmonics.  A
general formula, which may easily be verified, for such integrals is
$$(8\pi )^{-1}\oint n_A{}^{A'}n_B{}^{B'}\cdots n_C{}^{C'}(t\cdot n)^{-\nu -2}
  \d{\cal S}=2^\nu (\nu +1)^{-1}t_{(A}{}^{A'}t_B{}^{B'}\cdots
t_{C)}{}^{C'}(t^2)^{-\nu -1}\, ,\eek$$
where $\nu$ is the number of $n$'s.  (This equation holds for any timelike
vector $t^a$.)  The cases we shall need are, when $t^2=1$,
$$(8\pi )^{-1}\oint n_an_b (t\cdot n)^{-4}\d {\cal S}=(1/3)\bigl[ 4t_at_b-\eta
_{ab}\bigr]
  \eek$$
and
$$\deqn{
  (8\pi )^{-1}\oint n_an_bn_cn_d(t\cdot n)^{-6}\d {\cal S}
  = (2/45)
    \bigl[ \eta _{ab}\eta _{cd}+\eta _{ac}\eta _{bd}+\eta _{ad}\eta _{bc}\bigr]
\cr
\ind +\hbox{terms involving at least two of } t_a\, ,\ t_b\, ,\ t_c\, ,\ t_d\,
.\keq\cr}$$
(Here $\eta _{ab}$ is the Minkowski metric.)

The first integral in (\anot{}) extracts a combination of the $l=0$ and $l=1$
parts of $\delta \phi$ in the Bondi frame.  This is nothing but
$(8\pi /3)\bigl[ 4t_a\xi\cdot t-\xi _a\bigr]$,
where $\xi =\xi _an^a/(t\cdot n)^2$.  Similarly, the second integral extracts
a combination of the $l=0$ and $l=1$ parts of $\phi \delta t\cdot n$.  Here
$\delta t ^a$ must be purely $l=1$ (since $t^2=1$), and so only the $l=0$ and
$l=2$ parts of $\phi$ can contribute.  However, $\phi$ has no $l=0$ part.
Indeed, let us put
$$\phi =\phi _{ab}n^an^b(t\cdot n)^{-3}+\cdots\, ,\eek$$\xdef\phab{{\the\EEK}}%
where $\phi _{ab}$ is symmetric, trace--free and transverse to $t^a$.  Then
we have
$$\eqalign{
 (8\pi )^{-1}\oint n_a\phi (\delta t\cdot n)(t\cdot n)^{-3}\d{\cal S}
  &=(8\pi )^{-1}\phi _{bc}\delta t_{d}\oint n_an^bn^cn^d(t\cdot
n)^{-5}\d{\cal S}\cr
  &=(4/45)\phi _{ab}\delta t^b\cr}\eek$$
Therefore we have
$$\xi _a=-(4/5)\phi _{ab}\delta t^b\, .\eek$$

It remains to compute $\delta t^a$.  The formula for the Bondi--Sachs momentum
is usually written as
$$P_aW^a=-(4\pi G)^{-1}\oint \bigl[\psi _2+\sigma \dot{\overline\sigma}\bigr]
   W \d{\cal S}\eek$$
where $W$ has type $\{ 1,1\}$.  Putting $W=W_al^a$, we have
$$P_a=-(4\pi G)^{-1}\oint \bigl[\psi _2+\sigma \dot{\overline\sigma}\bigr]
   l_a \d{\cal S}\, .\eek$$
We need to compute the variation of the Bondi--Sachs momentum on $S$ as the
shea
is perturbed, given that all quantities are constant on $Z$.  From the
momentum--loss formula, we have
$$\delta P_a=(4\pi G)^{-1}\int\bigl[ \dot s\dot{\overline\sigma}
  +\dot{\overline s}\dot\sigma\bigr] l_a\d{\cal S}\d u\,
.\eek$$\xdef\Dip{\the\EEK}%
(The plus sign arises because $Z$ is to the future of $S$.)
As usual, we have $\delta t^a=m^{-1}\delta P^a-(\delta m)m^{-2}t^a$, where $m$
is the Bondi--Sachs mass.  The second term here gives no contribution to $\xi
_a$, since $\phi _{ab}$ is transverse to $t^a$.  Thus
$$\xi _a=-(4/5)m^{-1}\phi _{ab}\delta P^b\, ,\eek$$
with $\delta P_a$ as in (\Dip{}).

\subsection{Definition and Invariance of the Angular Momentum}

\noindent We shall be interested in the total angular momentum of the
space--time, as measured from $\scrip$.  This will be defined as a suitable
limit of the angular momentum along cuts tending to $u=-\infty$.  Once the
cut is taken so far in the past that $\sigma$ is independent of $u$, it turns
out that the angular momentum is independent of the cut,  when proper care is
taken to identify the spaces of origins relative to the cuts.  Here we show how
this comes about.

At any cut $S$, the space $\M (S)$ of origins for the twistorial angular
momentum is defined as follows.  Let $\sigma _S$ be the shear of the cut.  Then
$\M (S)$ is the set of those solutions to the equation $\eth ^2\lambda =\sigma
_S$ which have real $l=0$ and $l=1$ parts with respect to the metric defined by
the Bondi--Sachs momentum.  The solution with zero $l=0$ and $l=1$ parts is
denoted $\phi (S)$ and called the \it focus \rm of $S$.  The angular momentum
about $\lambda$ is given by
$$\eqalign{\mu _{A'B'}c^{A'B'}
  &=-(8\pi G)^{-1}\oint _S\bigl[ \psi _1c+(\psi _2+\sigma\dot{\overline\sigma}
    )(2c\eth\lambda -\lambda\eth c)\bigr] \d{\cal S}\cr
  &=-(8\pi G)^{-1}\oint _S\bigl[ \psi _1c+\psi _2(2c\eth\lambda -\lambda\eth
 c)\bigr] \d{\cal S}\cr}\eek$$
in our case.  Now consider a second cut, $\hat S$, supertranslated by
$\alpha$, in the same quiescent domain.  Then we have
$$\eqalign{\sigma _{\hat S}&=\sigma _S-\eth ^2\alpha\cr
  {\hat\psi}_1(\hat S,\theta ,\varphi )&=\psi _1(\hat S,\theta ,\varphi )+
    3(\eth\alpha )\psi _2(\hat S,\theta ,\varphi )\cr
  &=\psi _1(S,\theta ,\varphi )+\alpha\eth\psi _2(S,\theta ,\varphi )
    +3(\eth\alpha )\psi _2(\hat S,\theta ,\varphi )\cr
  {\hat\psi}_2 (\hat S,\theta ,\varphi )&=\psi _2(S,\theta ,\varphi )\cr
  \phi (\hat S)&=\phi (S)-\xi\cr}\eek$$
where $\xi$ is equal to $\alpha$ with its $l=0$ and $l=1$ parts (with respect
to
the frame defined by the Bondi--Sachs momentum) set to zero.
We identify the spaces of origins according to
$$\matrix{\M (S)&\to&\M (\hat S )\cr \lambda&\mapsto&\hat\lambda =
\lambda +\alpha\cr}\eek$$
Then we have
$$\deqn{
\mu _{A'B'}(\hat S,\phi (\hat S ) ) c^{A'B'}\cr
\ind\eql -(8\pi G)^{-1}\oint \d{\cal S}\bigl[ c\psi _1+c\alpha\eth\psi _2
    +3c(\eth\alpha )\psi _2 +\psi _2(2c\eth (\phi -\xi )
    -(\phi -\xi )\eth c
    \bigr]\cr
\ind\eql \mu _{A'B'}(S,\phi ( S ) ) c^{A'B'}
    -(8\pi G)^{-1}\oint \d{\cal S}\bigl[ c\alpha\eth\psi _2
    +3c(\eth\alpha )\psi _2 -\psi _2(2c\eth \xi     + \xi \eth c )
    \bigr]\cr
\ind\eql \mu _{A'B'}(S,\phi ( S ) ) c^{A'B'}
    -(8\pi G)^{-1}\oint \d{\cal S}\bigl[
    2c\eth (\alpha -\xi )\psi _2 -(\alpha -\xi )(\eth c)\psi _2
    \bigr]\cr
\ind\eql \mu _{A'B'}(S,\phi ( S ) +(\alpha -\xi )) c^{A'B'}\keq\cr}$$
We see that the angular momenta at the two cuts are related by the correct
change--of--origin formula, if the difference between the foci of the two cuts
is considered to be the projection of the supertranslation relating the cuts to
its $l=0$ and $l=1$ parts relative to the metric defined by the Bondi--Sachs
momentum.

To compute the total angular momentum, then, it will be no essential loss of
generality to choose the cut to be a translation of $Z$.  (The astute reader
will however anticipate that this leads to a somewhat anomalous behavior for
the
associated Hamiltonian vector fields.  This will be explained in Section 7.)
Thi
we shall do, and assume that they are both $u=$ constant cuts of a Bondi
coordinate system.

\section{Computation of the Vector Fields}

\noindent We now come to the problem of inverting the equations
$$\eqalign{\nabla P_a&=\varOmega (\cdot ,V_a)\cr
  \nabla \mu _{A'B'}&=\varOmega (\cdot ,V_{A'B'})\cr}\eek$$
to find the Hamiltonian vector fields $V_a$ and $V_{A'B'}$ for the momentum
and angular momentum.

It will be convenient to use a trick to avoid taking the limit $u\to -\infty$.
We describe it first in the case of the momentum (although most of the
advantage
comes in simplifying the angular momentum calculation).  The total momentum is
defined as
$$P_a^{\rm total}=\lim _{u\to -\infty} P_a(u)\, ,\eek$$
where $P_a(u)$ is the momentum at the cut $u$.  In fact, $P_a(u)$ is
independent of $u$ for negative enough $u$ (since the news is zero in a
neighborhood of $u=-\infty$).  We shall want to compute $\nabla P_a^{\rm
total}$, which is to say $\langle s,\nabla P_a^{\rm total}\rangle$ for all
admissible $s$.  Since each $\dot s$ has compact support, it is enough to
comput
$\langle s,\nabla P_a(u)\rangle$ for $u$ so negative that the news is zero and
also $\dot s$ is zero at and to the past of $u$.  In the calculations below,
thi
means we may assume that the support of $\dot s$ lies to the future of $u$;
this
has the effect of eliminating some boundary terms.  The same argument applies
to
angular momentum, and here the advantage is that it saves us from having to
compensate for the origin--dependence by introducing larger and larger
translations (so that the total angular momentum refers to a finite origin, not
one at $u= -\infty$).

In order to reserve the symbol $u$ for the parameter which will appear in the
vector field, it will be convenient to call the value of the Bondi parameter at
which the momentum and angular momentum are to be evaluated $u_0$.  The
parameter value of the cut $Z$ will be $Z_0$.  In keeping with the comments
above, we shall always have $u_0<u<Z_0$.
We consider, in effect, not quite the original function
space, but rather one in which the shear is constant for $u<u_0+\epsilon$ for
some $\epsilon >0$ (depending on the particular shear).

We may invert the equation
$$\lambda =\varOmega (\cdot ,V)\eek$$
as follows.  In the cases of interest, $\langle s,\lambda\rangle$ will be given
by a sum of integrals of smooth functions times derivatives of $s$ over the
region $u_0\leq u\leq Z_0$.
Let $\dang$ be the kernel of the
identity map on $C^\infty (S^2,\{ 3,-1\} )$.  Then one can verify
$$-4\pi G\varOmega\bigl( H(u-\cdot)\dang +H(u_0+0 -\cdot )\dang
  ,V^{(1,0)}\bigr)= V^{(1,0)}(u)\, ,\eek$$
where the integral for $\varOmega$ extends over the region $u_0\leq u\leq Z_0$,
the symbol $H$ represents the Heaviside step function, and $H(u_0+0-\cdot )$ is
short for writing $H(u_0+\epsilon -\cdot )$ and taking the limit
$\epsilon\downarrow 0$ at the end.  (The dot stands for the retarded time
parameter appearing in the integral for $\varOmega$.)
Therefore we have
$$V^{(1,0)}(u)=-4\pi G\langle H((u-\cdot)\dang +H(u_0+0 -\cdot )\dang ,\lambda
    \rangle\, ,\eek$$
and similarly for $V^{(0,1)}$, for which we use $\overline\dang$.

We start with the momentum,
$$P_aW^a=-(4\pi G)^{-1}\oint \bigl[\psi _2+\sigma \dot{\overline\sigma}\bigr]
   W \d{\cal S}\, .\eek$$
We computed above
$$\eqalign{
W^a\delta P_a&=(4\pi G)^{-1}W^a\int\bigl[ \dot s\dot{\overline\sigma}
  +\dot{\overline s}\dot\sigma\bigr] l_a\d{\cal S}\d u\cr
 &=-(4\pi G)^{-1}W^a\int\bigl[ s\ddot{\overline\sigma}
  +{\overline s}\ddot\sigma\bigr] l_a\d{\cal S}\d u\cr}\eek$$
Here the integration is over $\scrip$ from $u_0$ to $Z_0$, with the standard
orientation.  We find
$$V_a^{(1,0)}=\dot\sigma (u)\, l_a\, ,\eek$$
and of course the conjugate is
$$V_a^{(0,1)}=\dot{\overline\sigma} (u) \, l_a\, .\eek$$
This is as expected.

The computation for the angular momentum is lengthier.  Most of it is
straight\-forward, though, and we shall present only the main steps.
We recall
$$\eqalign{\mu
_{A'B'}c^{A'B'}&=-(8\pi G)^{-1}\oint \bigl[ \psi _1c
   +(\psi _2+\sigma \dot{\overline\sigma} )(2c\eth\phi -\phi\eth c)\bigr]
    \d{\cal S}\cr
  &=-(8\pi G)^{-1}\oint \bigl[ \psi _1c
   +\psi _2(2c\eth\phi -\phi\eth c)\bigr] \d{\cal S}\cr}\eek$$
at $u_0$, since $\dot\sigma =0$ there.  It will be convenient to write
$$\delta\mu ^{A'B'}c_{A'B'}=A+B+C\eek$$
where
$$\eqalign{A&=-(8\pi G)^{-1}\oint _{u_0}\delta\psi _1c\, \d{\cal S}\cr
  B&=-(8\pi G)^{-1}\oint _{u_0}(\delta \psi _2)
    (2c\eth\phi -\phi\eth c)\, \d{\cal S}\cr
  C&=-(8\pi G)^{-1}\oint _{u_0}\psi _2
    \bigl( 2c\eth\delta\phi -(\delta\phi )\eth c\bigr)\, \d{\cal S}\cr}\eek$$
We shall compute the contributions to the vector field from each of these terms
separately.

We begin with the contribution from $A$.  A straightforward lengthy calculation
gives
$$\eqalignno{
  A&=(8\pi G)^{-1}\oint\d{\cal S} \, c\eth\int \d w\, (w-u_0)H(w-u_0)
    \bigl[ s\ddot{\overline\sigma} +\sigma\ddot{\overline s}\bigr]\cr
  &\phantom{=} -2(8\pi G)^{-1}\oint\d {\cal S}\, c\int\d w\, H(w-u_0)
    \bigl[ s\eth\dot{\overline\sigma} +\sigma\dot{\overline
s}\bigr]&\keek\cr}$$
We find
$$V_A^{(1,0)}=c\eth\sigma (u)+(3/2)(\eth c)\sigma (u)
  +(1/2)(u-u_0)(\eth c)\dot\sigma (u) +c\eth\sigma (u_0)+(3/2)(\eth c)\sigma
(u_0)\, \eek$$
and
$$V_A^{(0,1)}=c\eth\overline\sigma (u)-(1/2)(\eth c)\overline\sigma (u)
  +(1/2)(u-u_0)(\eth c)\dot{\overline\sigma} (u)
  -c\eth\overline\sigma (u_0)+(1/2)(\eth c)\overline\sigma (u_0)\, .\eek$$
(Here the subscript $A$ is not a spinor index, but indicates these are the
vector fields derived from $A$.  Similar comments apply to $V_B$ and $V_C$,
below.)

For $B$, we find
$$B=(8\pi G)^{-1}\oint\d {\cal S} \bigl[ 2c\eth\phi (u_0)-\phi (u_0)\eth
c\bigr]
  \int\d w\, H(w-u_0)\bigl[ -\eth ^2\dot{\overline\sigma}
  +\dot s\dot{\overline\sigma} +\dot\sigma \dot{\overline s}\bigr]\eek$$
and
$$V_B^{(1,0)}=(1/2)\bigl[ 2c\eth\phi (u_0)-\phi (u_0)\eth c\bigr]
  \dot\sigma (u)-2c\eth\sigma (u_0)-3(\eth c)\sigma (u_0)\eek$$
and
$$V_B^{(0,1)}=(1/2)\bigl[ 2c\eth\phi (u_0)-\phi (u_0)\eth c\bigr]
  \dot{\overline\sigma}(u)\, .\eek$$

For $C$, we find
$$\eqalignno{
  C&=-(8\pi G)\oint _{u_0}\d {\cal S}\, \psi _2\bigl[ 2c\eth -(\eth c)\bigr]
      \eth ^{-2}s\cr
 &\phantom{=} -(8\pi G)^{-1}\left[ \oint _{u_0}\d {\cal S}\, \psi _2
    \bigl[ 2c\eth -(\eth c)\bigr] n^a(t\cdot n)^{-2}\right]\cr
 &\phantom{=}\times (-4/5)\phi _{ab}(4\pi Gm)^{-1}\oint\d{\cal S}
  \int\d w\, H(w-u_0)\bigl[ \dot s\dot{\overline\sigma}
   +\dot\sigma\dot{\overline s}\bigr] l^b\, .&\keek\cr}$$
Here $t^a$ is the unit vector in the direction of the Bondi--Sachs momentum at
$u_0$, and $m$ is the Bondi--Sachs mass at this cut.  The term in the large
square brackets can be computed:  it is
$$-8\pi Gmc^{A'}{}_{Q'}t^{AQ'}\, .\eek$$
Then we get
$$V_C^{(1,0)}=-(4/5)c^{A'}{}_{Q'}t^{AQ'}\phi _{ab}l^b\dot\sigma (u)\eek$$
and
$$V_C^{(0,1)}=-(4/5)c^{A'}{}_{Q'}t^{AQ'}\phi _{ab}l^b\dot{\overline\sigma}
(u)\, .\eek$$

Collecting our results, the vector field conjugate to this component of the
angular momentum is
$$V=V_A+V_B+V_C+\hbox{a possible gauge term.}\eek$$
The gauge term is not fixed by the mathematical structure so far imposed, and
must be determined by other considerations.  The most natural thing to do is to
try to eliminate the explicit dependence of the vector field on $\sigma (u_0)$.
This is possible.  For the terms which
involve $\sigma (u_0)$ are (from $V_A^{(1,0)}$, $V_A^{(0,1)}$ and
$V_B^{(1,0)}$)
$$-\bigl[ c\eth \sigma (u_0)+(3/2)(\eth c)\sigma
(u_0)\bigr]\eek$$\xdef\coop{{\the\EEK}}%
and
$$-\bigl[ c\eth\overline\sigma (u_0)-(1/2)(\eth c)\overline\sigma (u_0)\bigr]\,
,\eek$$\xdef\ccoop{{\the\EEK}}%
which precisely represent a Lorentz action on $\sigma (u_0)$, and so will be
purely electric contributions if $\sigma (u_0)$ is.

The vector field, then, is
$$\eqalignno{
V^{(1,0)}&=(1/2)(\eth c)\bigl[ (u-u_0)\dot\sigma (u)
          +3\sigma (u)\bigr]
          +c\eth \sigma (u)\cr
&\phantom{=}+(1/2)\bigl[ 2c\eth\phi (u_0)-\phi (u_0)\eth c\bigr]
          \dot\sigma (u)\cr
&\phantom{=}-(4/5)c^{A'}{}_{Q'}t^{AQ'}\phi _{ab}l^b\dot\sigma (u)\cr
V^{(0,1)}&=-(1/2)(\eth c)\bigl[ -(u-u_0)\dot{\overline\sigma}(u)
          +\overline\sigma (u)\bigr] +c\eth\overline\sigma (u)\cr
&\phantom{=}+(1/2)\bigl[ 2c\eth\phi (u_0)-\phi (u_0)\eth c\bigr]
  \dot{\overline\sigma} (u)\cr
&\phantom{=}-(4/5)c^{A'}{}_{Q'}t^{AQ'}\phi _{ab}l^b\dot{\overline\sigma}
(u)&\keek\cr}$$\xdef\hamvec{{\the\EEK}}%
where $\phi _{ab}$ is the quadrupole part of $\phi$ at $u_0$,
equation (\phab{}).

\section{Interpretation of the Vector Field}

\noindent The terms in the Hamiltonian vector field have the following
interpretations.  The first line in each of (\hamvec{}a,b) is the effect of a
BMS Lorentz rotation determined by $c$ about the cut $u=u_0$.  The second line
is a conjugation with a supertranslation by $\phi (u_0)$, and (therefore) the
first two lines have the combined effect of a BMS Lorentz rotation about the
shear--free cut which is the focus of $u=u_0$.  These are in some sense the
terms one expects.  The third line is rather a
surprise, and represents an anomalous translation.

One's first thought might be that the third term is clearly unphysical, since
it was argued above that the angular momentum is independent of the cut chosen
in the quiescent regime, when the origins are correctly identified, and yet
the associated vector field depends on the cut and indeed on the shear of the
cut.  However, a closer examination will show that the angular momentum's
cut--independence actually forces this complicated behavior on the vector
fields.
The reason is that in order to identify the spaces of origins for angular
momenta at two different cuts, it is necessary to project the $l=0$ and
$l=1$ components of the supertranslation relating the cuts, and this projection
is done \it with respect to \rm the frame defined by the Bondi--Sachs momentum.
When we take the derivative $\nabla\mu _{A'B'}$, the dependence of the
projection on the Bondi--Sachs momentum gives a term, and inverting this
gives the anomalous contribution to the Hamiltonian vector field.  This
behavior is inescapable if we want a definition of angular momentum which is
supertranslation--invariant in regimes with $\sigma$ purely electric and
independent of $u$.

We wish to underscore that these translations will arise on quite general
grounds and are not some essentially twistorial oddity.  Suppose, as is
conventional, we accept the definition of angular momentum by linkages at
shear--free cuts  (Tamburino and Winicour 1965, Winicour 1980, Geroch and
Winicour 1981).  In a regime in which the shear is independent of $u$ and
purely
electric, we could extend this definition to arbitrary cuts, by defining the
angular momentum about a cut $Z$ to be that about the focus of the cut.  This
would seem to be unobjectionable, and it coincides with the twistorial
definition in this case.  Thus if the accepted definition by linkages is
extended, in the most reasonable way, to purely electric cuts in quiescent
regimes, we will have the same anomalous translations.

The vector field is \it not \rm in general tangent to the leaves of the
foliation of the phase space.  This is to be expected, because the leaves are
not individually Lorentz--invariant.  Indeed, the vector field carries one
secto
to another in precisely the expected fashion:
$$\eqalign{
  [\! [V^{(1,0)}]\!] &=c\eth [\! [\sigma ]\!] +(3/2)(\eth c)[\! [\sigma ]\!]\cr
  [\! [V^{(0,1)}]\!] &=c\eth [\! [\overline\sigma ]\!] -(1/2)(\eth c)
    [\! [\overline\sigma ]\!]\, .\cr}\eek$$

\section{Summary and Discussion}

\noindent The aim of this paper was to derive motions of future null infinity
to which the components of the angular momentum of space--time, as defined by
Penrose, are conjugate.  To do this, we first defined a phase space, consisting
essentially of the Bondi shears.  This phase space was reduced by an action of
the quotient Supertranslations/Trans\-la\-tions.  The angular momentum was
shown to be well--defined on the phase space, and the equation
$\nabla\mu _{A'B'}=\varOmega (\cdot ,V_{A'B'})$ inverted to find the vector
fields.
The vector
fields turned out to be those generated by BMS Lorentz rotations about the
focus of the cut at which the angular momentum was computed, together with
anomalous translations by an amount
$$-(4/5)c_{A'B'}t_A{}^{B'}\phi ^{ad} +\hbox{conjugate}
  =(4/5)\bigl[ c_{A'B'}\epsilon _{AB}+{\overline c}_{AB}\epsilon _{A'B'}\bigr]
  t^{BB'}\phi ^{ad}\, .\eek$$\xdef\anom{{\the\EEK}}%
\it Note that these
anomalous are present even in stationary space--times, \rm when our
reduction of the BMS group is the accepted one.  Also \it these terms are
presen
even in the weak--field limit.  \rm  (However, Minkowski space is singular,
from
this point of view, because the direction $t^a$ is not determined.)

Thus we have achieved our goal.  This removes one of the objections that had
been raised to the twistorial definition of angular momentum:  that it seemed
to have no relation to real motions of space--time.  One the other hand, the
significance of the motions we have derived needs to be elucidated.  This
will be done in detail elsewhere, but we give here some tentative comments.

The space--times we have been studying are supposed to be idealizations which
model isolated radiating systems.  Consider, then, a finite number of isolated
radiating systems (stars, for short) in some space--time which is otherwise
clos
to flat.  Thus we assume there is a region around each star modeled by the sort
of space--time we have been considering, and that there is background Minkowski
metric which approximates the physical metric far away from any star.  In a
region far from the center of any star, but also far from the other stars, the
Minkowksi metric determines a family
of Bondi coordinate systems; in fact, a Poincar\'e group's worth
of Bondi systems.  In general, the $u=$ constant cuts of these systems will
have shear as $u\to -\infty$.

Now in ordinary special--relativistic mechanics, conservation of momentum would
imply that only the relative positions of the stars are significant,
conservation of the angular momentum would imply that only the relative
orientations and boosts of the stars are significant.  Our results suggest that
perhaps what enters is not the boost, but the boost supplemented by the
anomalous translation.  (The combination of
$c_{A'B'}$ and $t^a$ in (\anom{}) detects the boost, rather than the
rotational,
parts of $c_{A'B'}$ in the frame determined by the Bondi--Sachs momentum.)
This
might mean that we cannot infer the boosts of stars from Doppler measurements
unless we also have some bounds on their supertranslations relative to us.

For another possible effect, it may be that the
vector fields can be interpreted as angular momentum operators for a quantum
scattering theory of gravitational radiation.  Then it would seem that
measurements of centers--of--mass of isolated gravitational systems couple to
th
shear of the cut at which the measurement is made.

All of the physics and mathematics in this paper has been conventional,
except for the reduction of the phase space.  It is there that a new physical
idea was introduced:  that we should seek a reduction by the quotient
Supertranslations/Trans\-la\-tions.  What is the evidence that the reduction we
have introduced is correct?

Certainly some reduction is necessary, for with too large a class of test
functions $s$, we cannot invert the equation $\langle s,\nabla \mu
_{A'B'}\rangle =\varOmega (s,V_{A'B'})$.  For stationary space--times, the
reduction we have used would presumably be accepted, since it is then a
reduction by active BMS motions.  In these cases, the anomalous contributions
will be present, and so one cannot criticize the reduction on the grounds that
anomalous contributions arise.  The real novelty of the reduction is for
non--stationary space--times.  In this
case, the reduction adopted is \it not \rm an action by the  asymptotic
isometry group.  Given that anomalies \it will \rm be
present, the relative simplicity of the results is circumstantial evidence,
whose force will be left to the judgement of the reader, that the reduction is
correct.

\references

\refbk{Arnowitt R L, Deser, S and Misner C W 1962}{Gravitation --- An
Introduction to Current Research}{ed L. Witten (New York:  Wiley) pp 227--65}

\refjl{Ashtekar A and Hansen R O 1978}{\JMP}{19}{1542--66}

\refjl{Ashtekar A and Magnon--Ashtekar A 1979}{\PRL}{43}{181--4}

\refjl{Ashtekar A and Streubel M 1981}{\PRS}{A376}{585--607}

\refjl{Bergmann P G and Goldberg I 1955}{Phys. Rev.}{98}{531--8}

\refjl{Bondi H 1960}{Nature}{186}{535}

\refjl{Bondi H, van der Burg M G J and Metzner A W K
1962}{\PRS}{A269}{21--52}

\refbk{Chernoff P R and Marsden J E 1974}{Properties of Infinite
Dimensional Hamiltonian Systems}{(Berlin--Heidelberg--New~York:  Springer)}

\refbk{Dirac P A M 1964}{Lectures on Quantum Mechanics}{(New York:  Belfer
Graduate School of Science, Yeshiva University}

\refjl{Dougan A J and Mason L J (1991)}{\PRL}{67}{2119--22}

\refjl{Friedrich H 1983}{Commun. Math. Phys.}{91}{445--72}

\refjl{Geroch R and Winicour J 1981}{\JMP}{22}{803--12}

\refjl{Helfer A D 1990}{\PL}{A150}{342--4}

\refjl{\dash 1992}{\CQG}{9}{1001-8}

\refbk{\dash 1993}{A Phase Space for Gravitational Radiation}{University of
Missouri preprint}

\refjl{Marsden J and Weinstein A 1974}{Rep. Math. Phys.}{5}{121--30}

\refjl{Moncrieff V 1975}{\JMP}{16}{493--8}

\refjl{\dash 1976}{\JMP}{17}{1893--1902}

\refjl{Newman E T and Penrose R 1966}{\JMP}{7}{863--70}

\refjl{Penrose R 1982}{\PRS}{A381}{53--63}

\refbk{Penrose R and Rindler W 1984--6}{Spinors and Space--Time}{Cambridge:
University Press}

\refjl{Rendall A D 1990}{\PRS}{A427}{221--39}

\refbk{\dash 1992}{Relativity Today}{ed Z. Perj\'es (Commack, New York:  Nova
Science) pp 57--64}

\refjl{Sachs R K 1962a}{\PRS}{A270}{103--26}

\refjl{\dash 1962b}{Phys. Rev.}{128}{2851--64}

\refjl{Shaw W T 1983}{\PRS}{A390}{191--215}

\refjl{Sommers P 1978}{\JMP}{19}{549--54}

\refjl{Tamburino L and Winicour J 1965}{\PRL}{15}{601--5}

\refjl{Tod K P 1983}{\PRS}{A388}{457--77}

\refjl{\dash 1984}{Twistor News.}{18}{3--6}

\refjl{\dash 1986}{\CQG}{3}{1169--89}

\refbk{Winicour J 1980}{General Relativity and Gravitation:  One Hundred
Years After the Birth of Albert Einstein}{ed A Held (New York:  Plenum) pp
71--96}

\bye